\documentclass[sn-apa]{sn-jnl}

\usepackage{graphicx}%
\usepackage{multirow}%
\usepackage{amsmath,amssymb,amsfonts}%
\usepackage{amsthm}%
\usepackage{mathrsfs}%
\usepackage[title]{appendix}%
\usepackage{xcolor}%
\usepackage{textcomp}%
\usepackage{manyfoot}%
\usepackage{booktabs}%
\usepackage{algorithm}%
\usepackage{algorithmicx}%
\usepackage{algpseudocode}%
\usepackage{listings}%
\usepackage{subfigure}%
\usepackage{caption}

\begin{document}

\title[Article Title]{Surprising Performances of Students with Autism in Classroom with NAO Robot}


\author[1]{\fnm{Qin} \sur{Yang}}\email{outskyyq@163.com}

\author[3]{\fnm{Huan} \sur{Lu}}\email{415953174@qq.com}
\equalcont{These authors contributed equally to this work.}

\author[4]{\fnm{Dandan} \sur{Liang}}\email{03275@njnu.edu.cn}
\equalcont{These authors contributed equally to this work.}

\author[2]{\fnm{Shengrong} \sur{Gong}}\email{shrgong@cslg.edu.cn}
\equalcont{These authors contributed equally to this work.}

\author*[2]{\fnm{Huanghao} \sur{Feng}}\email{fenghuanghao@cslg.edu.cn}
\equalcont{These authors contributed equally to this work.}

\affil[1]{\orgdiv{School of Computer and Information Technology}, \orgname{Northeast Petroleum University}, \orgaddress{\street{No.99 Xuefu Street}, \city{Daqing}, \postcode{163318}, \state{Heilongjiang}, \country{China}}}

\affil*[2]{\orgdiv{School of Computer Science and Engineering}, \orgname{Changshu Institute of Technology}, \orgaddress{\street{No. 99, South Third Ring Road}, \city{ChangShu}, \postcode{215500}, \state{Jiangsu}, \country{China}}}

\affil[3]{\orgname{Changshu Special Education School}, \orgaddress{\street{No.50, Songshan Road}, \city{ChangShu}, \postcode{215500}, \state{Jiangsu}, \country{China}}}

\affil[4]{\orgdiv{School of Chinese Language and Literature}, \orgname{Nanjing Normal University}, \orgaddress{\street{122 Ninghai Rd, Gulou District}, \city{Nanjing}, \postcode{210097}, \state{Jiangsu}, \country{China}}}


\abstract{Autism is a developmental disorder that manifests in early childhood and persists throughout life, profoundly affecting social behavior and hindering the acquisition of learning and social skills in those diagnosed. As technological advancements progress, an increasing array of technologies is being utilized to support the education of students with Autism Spectrum Disorder (ASD), aiming to improve their educational outcomes and social capabilities. Numerous studies on autism intervention have highlighted the effectiveness of social robots in behavioral treatments. However, research on the integration of social robots into classroom settings for children with autism remains sparse. This paper describes the design and implementation of a group experiment in a collective classroom setting mediated by the NAO robot. The experiment involved special education teachers and the NAO robot collaboratively conducting classroom activities, aiming to foster a dynamic learning environment through interactions among teachers, the robot, and students. Conducted in a special education school, this experiment served as a foundational study in anticipation of extended robot-assisted classroom sessions. Data from the experiment suggest that ASD students in classrooms equipped with the NAO robot exhibited notably better performance compared to those in regular classrooms. The humanoid features and body language of the NAO robot captivated the students' attention, particularly during talent shows and command tasks, where students demonstrated heightened engagement and a decrease in stereotypical repetitive behaviors and irrelevant minor movements commonly observed in regular settings. Our preliminary findings indicate that the NAO robot significantly enhances focus and classroom engagement among students with ASD, potentially improving educational performance and fostering better social behaviors.}




\maketitle

\section{Introduction}\label{sec1}

Autism Spectrum Disorder (ASD), also known as autism, comprises a range of neurodevelopmental disorders that primarily manifest through social communication difficulties, restricted interests or activities, and repetitive behaviors \cite{ref1}. The incidence of ASD is increasing annually. According to the latest 2020 statistical data from the Centers for Disease Control and Prevention (CDC), in the United States, 1 in every 36 children aged 8 is diagnosed with ASD, with the prevalence in boys being approximately four times higher than in girls. In recent years, propelled by the rapid advancement of the intelligent Internet of Things and digitalization, social robots have found increasing application across various scenarios. Research in typical children's classrooms has demonstrated the potential for students and robots to co-learn effectively in real group settings \cite{ref3}\cite{ref4}\cite{ref5}. As suggested by educational theorists \cite{ref6}, robot-assisted activities hold significant promise for enhancing classroom teaching, particularly as children actively engage with constructs in their external environment, thereby facilitating more effective learning. Currently, social robots typically fulfill three roles in educational settings: A) Teacher; B) Teacher’s Assistant; and C) Student’s Companion.

Children with autism, who often exhibit specific impairments in facial recognition leading to deficits in social interaction, tend to focus more on inanimate objects \cite{ref7}. This predisposition makes them more receptive to initial social interactions with robots \cite{ref115}, turning social robots into a valuable assistive tool in ASD interventions \cite{ref9}. However, most existing studies on the interactions between children with autism and social robots are limited to one-on-one settings, typically within isolated spaces at rehabilitation centers. This approach has primarily fostered individualized interventions, which do not equip ASD students with the necessary group interaction skills crucial for effective participation in mainstream classroom environments \cite{ref10}\cite{ref11}\cite{ref12}. Group interaction skills are pivotal, influencing various learning behaviors in classroom settings \cite{ref13}. Drawing from the insights of prior research, this study designs an experimental framework for integrating social robots into classrooms specifically tailored for children with autism \cite{ref14}, thus paving the way for a robot-assisted classroom model for ASD. Our research aims to assess the impact of the NAO robot on classroom performance among children with autism.

In this paper, we present the preliminary findings of our study. As a pilot investigation, we observed the classroom performance of children with autism in both robot-assisted and regular classroom settings. Collaborating with special education teachers, we developed a robot-assisted cooperative curriculum based on long-term observations of behavior in regular classrooms. Using video content annotated from multiple perspectives, we analyzed and compared the classroom performance in both settings. This analysis encompassed four dimensions: classroom attention, classroom communication, classroom interaction assessment, and classroom emotion, to evaluate the effects of the NAO robot on children with autism.

The remainder of this paper is organized as follows: Section 2 discusses the related work. Section 3 details the participants, research hypotheses, curriculum design, experimental setup, and the coding scheme used. Section 4 presents the analysis of the experimental data. The conclusions are outlined in Section 5.

\section{Related works}\label{sec2}
\subsection{Research on NAO Robot in the Field of ASD}\label{subsec2}

Numerous studies have explored the interaction between social robots and children with autism. \cite{ref15} highlighted that the simple and predictable humanoid appearance of social robots facilitates social engagement among children with autism and offers novel sensory stimuli. A study by \cite{ref16} employed the NAO robot to engage children with autism in interactive games, observing that children maintained attention on the NAO robot for over 50$\%$ of the activity time. Furthermore, it was noted that the children made more eye contact with the robot and exhibited fewer attention shifts when the NAO robot spoke \cite{ref17}. \cite{ref18} suggested that children with autism might prefer interacting with robots over human teachers, as robots occupy a unique niche between toys and humans.

Attention perception, particularly in individuals with ASD, is a complex process \cite{ref19}. These individuals often display atypical behavioral patterns \cite{ref20}. A study by \cite{ref21} on the visual attention of children with autism found that, compared to typically developing children, those with ASD were less likely to use their eyes to seek visual cues and more inclined to adjust their visual attention by turning their heads \cite{ref22}. Research by \cite{ref23} noted that among commercially available social robots, the NAO robot is most frequently discussed for its capabilities in remote control, semi-autonomous, and fully autonomous operations. A comparative experiment by \cite{ref24} on gesture intervention using two NAO robots showed that robot intervention was as effective as human intervention in helping children with autism learn gesture recognition and skills, with children in the robot-based group more likely to establish eye contact with the robot.

\cite{ref25} used a multiple baseline design for an ABA-based intervention mediated by the NAO robot, assisting children with autism in answering 'wh' questions—a fundamental component of daily verbal interaction and social behavior. The findings indicated that social robots could effectively enhance the language and communication skills of individuals with autism, with efficacy comparable to human therapists \cite{ref26}.

Although it is generally believed that children’s attention and engagement with robots diminish over time, children with autism may maintain sustained attention on robots \cite{ref27}. In Pivotal Response Treatment (PRT) therapy sessions using the NAO robot, \cite{ref28} found that children with autism's attention and engagement levels remained consistent throughout the activity. \cite{ref29} utilized the NAO robot for one-on-one social skill intervention training at a rehabilitation center, discovering that children with autism stayed engaged across multiple interactive sessions over an extended period.

Current research predominantly focuses on one-on-one interventions with social robots for children with autism. Given that school is a critical period in children’s lives, experiencing growth in an inclusive group environment can mitigate problematic behaviors and actively foster the development of children and adolescents \cite{ref30}. A pilot study by \cite{ref12} introduced the robot Pepper into classrooms for children with autism to teach them necessary group interaction skills. This study demonstrated the feasibility of deploying social robots in actual classroom settings, offering a promising alternative for children with special needs.

\subsection{Related Theoretical Framework}\label{subsec2}

Constructivism, a cognitive psychology branch also known as structuralism, highlights "context," "collaboration," "conversation," and "meaning construction" as key elements in the learning environment. It promotes student-centered learning, emphasizing active exploration, discovery, and knowledge construction. With technological advancements, constructivism has increasingly influenced teaching practices and educational reforms, particularly in special education. It recognizes the importance of using technology to create learning situations that enhance communication and education for ASD students, fostering their knowledge and skills acquisition \cite{ref32}.

On the other hand, Behaviorism Learning Theory, often referred to as "stimulus-response" theory, views learning as a linkage between stimulus and response, regarding learners as a "black box" where observable behavior is the primary focus. Applied Behavior Analysis (ABA) and the Structured Teaching strategy (TEACCH), pioneered by Professor Eric Schopler, are key behaviorism-based interventions. These approaches have proven effective in improving understanding and emotional stability in children with ASD \cite{ref37} \cite{ref38}.

However, behaviorism often addresses only superficial behavioral aspects, whereas constructivism-based interventions like the “Big Social” system focus on holistic development, including cognitive and social skills. Such constructivist approaches leverage group dynamics and observational learning to enhance joint attention and reduce stereotypical behaviors in children with ASD \cite{ref40} \cite{ref41}. As constructivism and technology continue to evolve, group-based social exploration is becoming increasingly relevant in ASD education, promoting initiative, cooperation, and creativity \cite{ref42}.

\section{Methods}\label{sec3}

This study adopted a mixed-methods approach, integrating both quantitative and qualitative techniques to gather and analyze data, including video analysis. The conclusions were derived from the outcomes of these analyses.

\subsection{Participants}\label{subsec3}

Participants were recruited from special education schools in S city. All participants had been diagnosed with autism by independent agencies not affiliated with this study. The inclusion criteria for the study were as follows: 1) Diagnosis of autism; 2) Age between 9 and 11 years; 3) Absence of auditory or visual impairments; 4) Ability to understand simple instructions; 5) Absence of aggressive or other severe behavioral problems. 

After screening, a total of 6 children participated in the experiment(see Table 1)

\begin{table}[h]
\caption{Paricipants information}\label{tab1}%
\begin{tabular}{@{}llllll@{}}
\toprule
ID & Age  & Sex  & CARS  & WISC & Verbal\\
\midrule
SN001    & 11   & M    & 42  & 87 & \checkmark \\
SN002    & 9    & M    & 47  & 73 & - \\
SN003    & 10   & F    & 37  & 90 & \checkmark \\
SN004    & 9    & M    & 39  & 89 & \checkmark \\
SN005    & 9    & M    & 40  & 85 & \checkmark \\
SN005    & 10   & M    & 41  & 81 & \checkmark \\
\botrule
\end{tabular}
\end{table}

Before the study began, special education teachers, experts, and parents of all participants had signed written informed consent forms and agreements to use the video data for research purposes. Participants could withdraw from the study at any time if they wished to discontinue or opt-out. All experimental lessons were individually rehearsed with special education teachers before starting to enhance classroom teaching fluidity and develop teacher’s skills in human-robot collaboration. 

\subsection{Research Hypotheses}\label{subsec3}
A social robot NAO was introduced into ASD children’s classrooms to promote courses teaching through interactions between robot-teacher, robot-students, and to enliven the classroom situation. The research hypotheses are as follows:

\begin{enumerate}[1)]
\item ASD students in the robot-assisted classroom will have higher online attention than in a regular classroom.

\item ASD students in the robot-assisted classroom will engage in more classroom-related communication behaviors.

\item ASD students in the robot-assisted classroom will receive higher classroom activity evaluation scores than in the regular classroom.

\item ASD students in the robot-assisted classroom will exhibit better emotional states.
\end{enumerate}

\subsection{ Curriculum Design}\label{subsec3}

The curriculum design is carried out according to the following requirements:
\begin{enumerate}[1)]
\item	Teaching Factors Analysis

Teaching analysis and structured teaching are used to analyze the learning characteristics of children with autism, curriculum teaching objectives, and curriculum content. The teaching design of social robot is combined with special education teachers.

\item   Teaching Adaptability Analysis

Filter out parts of the curriculum suitable for robot-assisted teaching. If inappropriate, present them with animations and slideshows.

\item	Structured Instructional Design

After selecting the teaching content of the course, the content is restructured to be more in line with the perceptual and cognitive characteristics of ASD children, making classroom teaching tasks clearer and classroom knowledge content easier to understand.

\end{enumerate}
All experimental courses have been discussed and confirmed with special education teachers. In the design process, around the classroom theme, knowledge points are presented step by step, combining constructivism learning theory to reduce the cognitive load brought by new knowledge and new objects. After all, by using Choregraphe software (see Fig 1), the course content has been designed as an experimental program. Before and after the formal course, a questionnaire survey will be conducted to test the children’s acquisition of classroom knowledge and collect feedback from teachers. This questionnaire is co-designed by special education teachers and the research team, exploring from multiple perspectives the impact of NAO on children and teachers in autism classroom scenario, and accumulating experience for subsequent experiments.

\begin{figure}[htbp]
	\centering
	\includegraphics[width=0.8\textwidth]{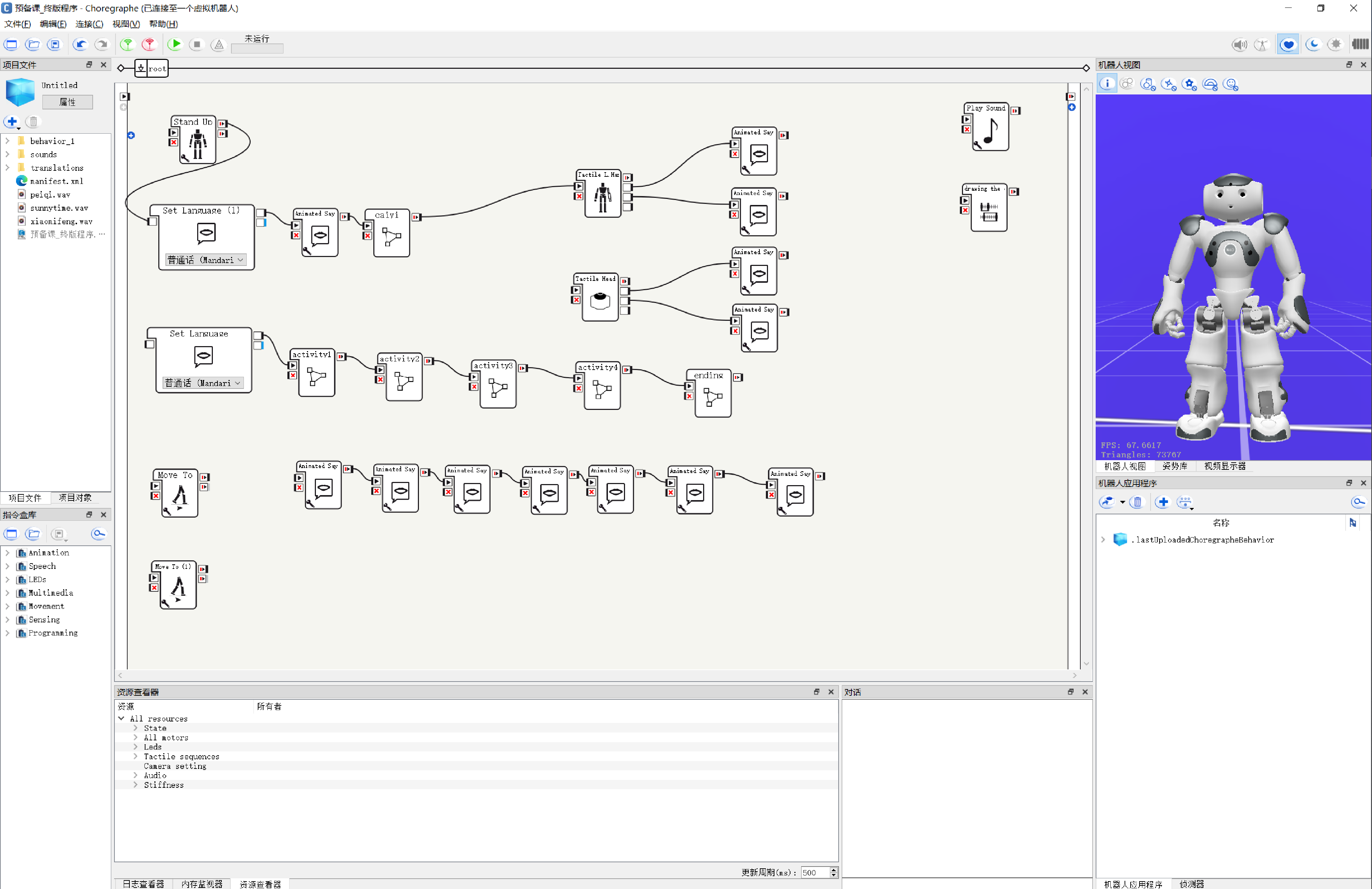}
	\caption{Choregraphe software}
	\label{fig1}
\end{figure}

\subsection{Experimental Design}\label{subsec3}
\subsubsection{NAO Robot}\label{subsubsec2}
The social robot NAO has used for robot-assisted classroom teaching (see Fig 2. (a)). NAO is an intelligent robot developed by SoftBank. It is capable of voice, and movement functions and can express a range of basic emotions such as anger, fear, and sadness. It can also infer emotional changes by learning the body language and facial expressions of its interactive partners.

\begin{figure}[htbp]
\centering
\subfigure[NAO Robot]
{
 	\begin{minipage}[b]{.4\linewidth}
        \centering
        \includegraphics[scale=1]{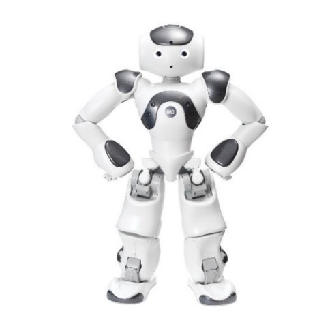}
    \end{minipage}
}
\subfigure[NAO components]
{
 	\begin{minipage}[b]{.4\linewidth}
        \centering
        \includegraphics[scale=1]{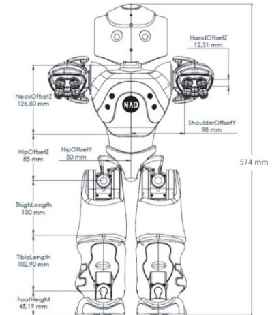}
    \end{minipage}
}
\caption{How a NAO robot looks like.}
\end{figure}

NAO is a programmable humanoid robot with a height of 57.4 centimeters. Its key components include:
\begin{enumerate}[1)]
\item	25 degrees of freedom (DOF).

\item	A series of sensors: 2 high-definition cameras, 4 microphones, two sets of ultrasonic distance sensors, 2 infrared emitters and receivers, 1 set of inertial sensing units (two gyroscopes, one accelerometer), 9 tactile sensors, and 8 pressure sensors.

\item	Devices for self-expression: a voice synthesizer, LED lights, and 2 high-quality speakers.

\item	CPU (ATOM E3845) with 1.91 GHz main frequency, 4GB DDR3 memory, 32GB SSD.
\item   2.9 Ah battery, providing 1.5 hours or more of battery life for NAO, depending on usage.
\end{enumerate}

\subsubsection{Classroom Environment}\label{subsubsec3}

\begin{figure}[htbp]
\centering
\includegraphics[width=0.9\textwidth]{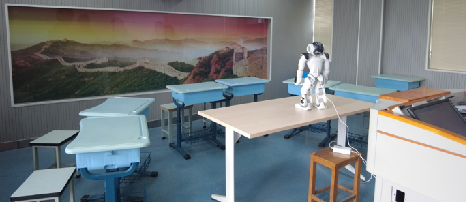}
\centering
\caption{The classroom environment in the experiment}\label{fig3}
\end{figure}

The experimental classroom has six seats arranged in two groups, forming a semi-circle facing the podium (see Fig 3). Some activities are conducted on a group basis. Observers and staff operating the robot are in an observation room at the back of the classroom, constructed of one-way mirror for sufficient safety and privacy. NAO was positioned in the center of the classroom, allowing children to observe it from different angles and facilitating face-to-face communication and interaction in various directions during the class.

\subsubsection{Camera Team}\label{subsubsec3}

\begin{figure}[htbp]
\centering
\includegraphics[width=0.9\textwidth]{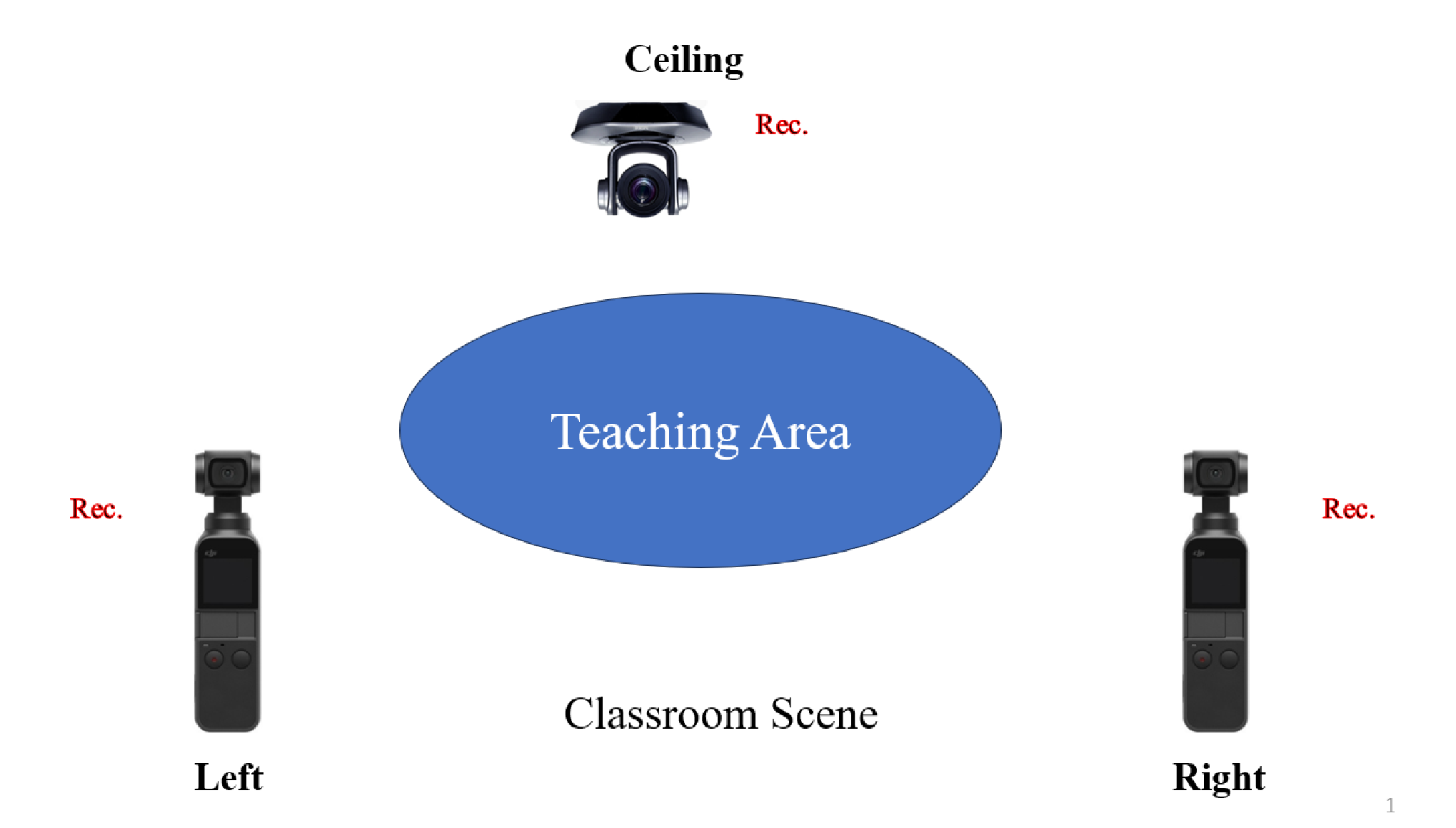}
\centering
\caption{The camera team set up in the experiment}\label{fig4}
\end{figure}

During the course, three cameras continuously record from different positions: left, right, and the classroom’s built-in camera, mounted on the ceiling above the blackboard (see Fig 4).
\emph{Camera 1 \& 2: DJI Pocket 2}.

The DJI Pocket 2, a lightweight compact camera weighing just 117 grams, features a 1/1.7” 64MP CMOS sensor, 8x zoom, and a 93° field of view. It supports video recording up to 4K/60fps and 64x slow motion. During the experiment, observers monitored the classroom in real-time using the DJI Mimo app on their mobile phones, capturing both rear and front angles. The camera's high resolution and frame rate are optimized for post-production in Final Cut Pro software.


The special education school classroom is equipped with a binocular intelligent tracking camera featuring a dual-lens system for automated tracking and recording. It includes a 1/2.8-inch CMOS sensor that produces full HD 1920*1080P video. The main tracking camera combines 20x optical and 10x digital zoom, while the positioning camera uses a 2.8mm focal length with a 2.1MP fixed-focus HD lens.

\subsubsection{Coding Software}\label{subsubsec3}
\begin{figure}[htbp]
\centering
\includegraphics[width=0.9\textwidth]{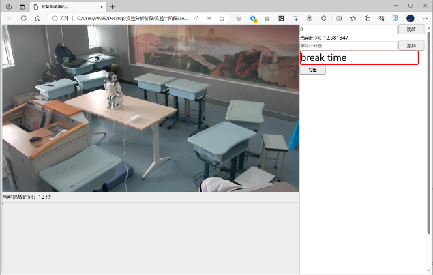}
\centering
\caption{Attention and Performance coding}\label{fig5}
\end{figure}

To enhance video observation, specialized video analysis software was developed (see Fig 5). It features an adaptive resolution video observation box on the left, where play and pause functions have been customized by modifying ASCII codes to improve usability. On the right, a "notebook" style video recording box allows observers to label observations when the video is paused. This software supports precise jumping and records the duration of specific behaviors, such as "break time," with an automatic timer that starts and stops via button clicks, and calculates duration for data analysis.

During the emotion observation phase, the video is manually advanced second by second using custom ASCII codes. Effective keys are set—"p" for positive, "e" for neutral, "n" for negative—and arrow keys adjust video progress. Pressing any key plays the video for one second and then pauses, allowing for detailed observation and labeling of emotions.

In the data processing phase, entries for attention are filtered and recorded based on duration criteria (2, 3, 5, or 10 seconds). This helps locate these durations in the original video for analysis. For behavior, data from regular and robot-assisted classrooms are summarized in one table, facilitating direct comparison and statistical analysis. For emotions, ASCII labels ("p" for positive, "n" for negative) are replaced and marked in color in Excel, enabling the selection and analysis of emotional instances. Microsoft Excel supports most of the basic analysis, streamlining the research on classroom performance.

\subsection{Implementation of the Experiment}\label{subsec3}
\begin{figure}[htbp]
\centering
\includegraphics[width=0.9\textwidth]{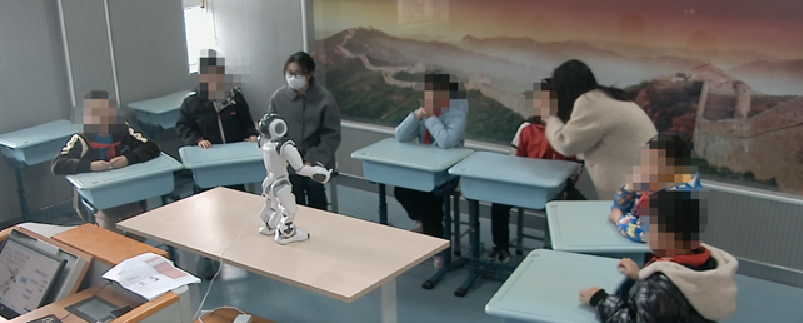}
\centering
\caption{Preparatory lesson for robot-assisted courses}\label{fig6}
\end{figure}

The special education school served as the experimental site for this study (see Fig 6). The children with autism had approximately three years of learning experience in the school. To eliminate some unknown factors, a group interaction preparatory lesson was co-designed with special education teachers before the series of robot-assisted courses. The preparatory lesson aimed to provide children with the basic skills required for participating in robot-assisted courses and laying the foundation for subsequent human-robot collaborative courses. An assistant teacher is assigned to a robot-assisted classroom in order to support instruction and prevent emergencies if necessary.

The activity design included: roll call, command listening, picture description, and group interaction. The course interspersed interesting group activities like small games and performances to teach children group interaction skills, including but not limited to following individual-specific instructions, group-specific instructions, distinguishing instructions for other groups, etc. With each activity described as follows:

\begin{enumerate}[1)]
\item	Roll Call

NAO introduces itself: “Hello everyone, I am your new classmate. My name is ‘NAO’, and I hope to become good friends with you! Now, I will do the roll call according to the list given by the teacher. If you hear your name, you can raise your hand or stand up, so I can get to know you.” After speaking, the robot begins the roll call and says, “Okay, let me first get to know the teaching assistant!” The teaching assistant raises her hand and stands up as an example for the children, then the robot starts calling names in sequence.

\item   Command Listening

This activity contains three commands: 1): raise your right arm and draw a circle in front of you, 2): clap your hands in front of you, and 3): shake hands with the teaching assistant in turn.

\item   Picture Description

Five pictures are displayed on the slideshow, with three segments: 1): What animal is in picture number two? 2): Which picture shows the rabbit? 3): NAO makes a “croak” frog sound, then asks the children to guess the animal and come to the screen to circle the picture of the frog.

\item   Group Interaction

In this segment, two adjacent children form a group. Before the activity starts, NAO and the teacher explain the rules to the children. The interaction segment consists of three parts: 1): the first group of students (SN001, SN002) stand up, 2): the second group of students (SN004) introduces the name of their neighbor (SN003), 3): the third group of students (SN005, SN006) face each other and shake hands.
\end{enumerate}

\subsection{Coding scheme}\label{subsec3}

Video coding was conducted in three rounds. The first round involved observing and labeling all behaviors and their audiences of one child per viewing, one time for each child. The second round focused solely on observing and labeling the pupil or head movements of one child per viewing to measure attention direction, one time for each child. The third round entailed observing and labeling the emotional state of one child per viewing, one time for each child. Children’s classroom performance included classroom attention, communication, interaction assessment, and emotion.

Attention: classroom attention was measured by observing children’s gaze and head direction during class. Three targets of attention direction were defined: 1) Teacher or Robot, 2) Blackboard or Screen, and 3) Classmate or Other places. If children’s attention was moved towards either of the first two targets (1 and 2), it was labeled as online. Offline was classified when participants did not focus on the first two targets. The percentage of online time during class was calculated by dividing the duration of the online target by the total class duration.

Communication: Classroom communication was measured by observing the frequency and target audience of all verbal and non-verbal communication during class. Such as raising a hand or pointing to a picture on the screen, answering what the picture is, requesting the teacher to add stars for oneself, and any sounds with clear words. Incomprehensible non-verbal sounds, verbal tics, etc., were not considered as proper communication. The target audience was categorized into four groups: 1) Teacher, 2) Robot, 3) Self, and 4) Classmate.

Interaction Assessment: Classroom interaction assessment was measured by children’s correct responses to themselves or their group. Comprising questions and activities from teachers and robot NAO, including responding when asked, staying quiet when others are asked, and responding when others or other groups are asked. Scoring was based on a three-tier system, with each tier worth 1 point: first tier) whether attention was turned to the questioner when NAO or the teacher asked a question; second tier) whether children responded after NAO or the teacher asked a question, including raising hands, standing up, or verbal responses; and finally third tier), whether children’s answers were correct. If the answer was correct after the teacher’s reminder, the score for that instance was multiplied by 0.5; incorrect or no responses scored zero.

$$
\mathrm{P}=\frac{R s+R b+R c+R c a \times 0.5}{3 n} \times 100 \%
$$

P is defined as the interaction assessment score, Rs is the number of times looking towards the questioner when asked, Rb is the number of times children responded after the question was asked, Rc is the number of times children independently and correctly answered, Rca is the number of times children correctly answered after the teacher’s prompt, 3 is the scoring level, and n is the total number of questions.

Emotion: Positive and negative class emotional statements were measured by observation. They were divided into three categories, each with several subcategories, determined collectively by observers. These were: Positive: (conscious smiling, raising hand to answer questions, attempting to touch the teacher and robot); Negative: (crying, sobbing, protesting, hitting oneself or others, throwing objects); Neutral: (if none of the above, classified as neutral). In emotion observation, some emotions require a comprehensive judgment based on surrounding clips.

Moreover, video coding adhered to the Interobserver Agreement (IOA) principle, with all videos independently coded by two professionally trained observers. The final IOA consistencies in this study were: Attention: 82.93\%, Communication: 83.84\%, Assessment: 97.56\%, Emotion: 92.78\%.

 \section{Data Analysis \& Discussion}\label{sec4}
 Before the implementation of the robot-assisted courses, long-term observations were made in the subjects’ regular classrooms, with classes randomly selected for comparative analysis.

\subsection{Comparative Analysis of Classroom Attention Data for ASD Children in Both Types of Classrooms}\label{subsec4}

Comparing the classroom attention data (see Fig 7), we found that the majority of ASD children spent a greater proportion of online time in robot-assisted classroom than in regular classroom (data in Table 2), with only SN003 slightly lower than in regular classroom. The proportion of time spent “looking at the teacher” in a robot-assisted classroom was less than that in a regular classroom (see Table 3). We believe that the inclusion of NAO goes a long way to capturing children's attention and increasing their potential motivation to participate in the classroom, thus reducing the focus on the teacher. 

\begin{figure}[htbp]
\centering
\includegraphics[width=0.9\textwidth]{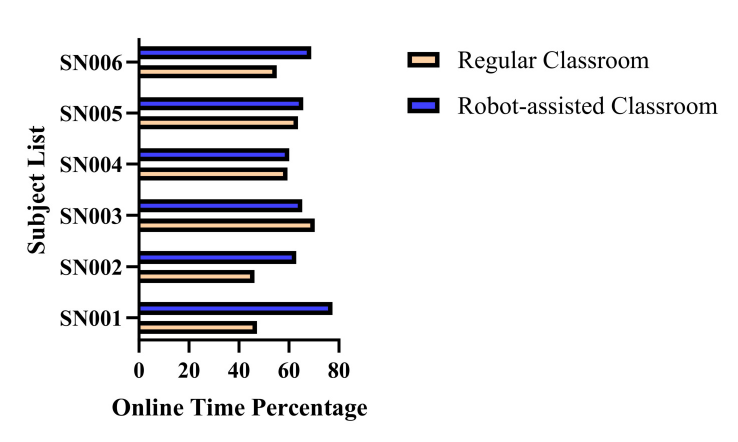}
\centering
\caption{Comparative Bar Chart of Subjects’Classroom Attention in Two Classrooms}\label{fig7}
\end{figure}

\begin{table}[b]
\caption{Proportion of Online Time of ASD Children in Two Classrooms}\label{tab2}%
\begin{tabular}{@{}lllll@{}}
\toprule
Attention & Regular Classroom  & Robot-assisted Classroom & Statement\\
\midrule
SN001    & 47.26$\%$   & 76.48$\%$    & $\uparrow$ \\
SN002    & 46.30$\%$   & 62.91$\%$    & $\uparrow$ \\
SN003    & 70.37$\%$   & 65.38$\%$    & $\downarrow$ \\
SN004    & 59.40$\%$   & 60.16$\%$    & $\uparrow$ \\
SN005    & 63.62$\%$   & 65.72$\%$    & $\uparrow$ \\
SN006    & 55.11       & 68.90$\%$    & $\uparrow$ \\
\botrule
\end{tabular}
\end{table}

\begin{table}[b]
\caption{Proportion of Time Spent “Looking at the Teacher” by ASD Children in Two Classrooms}\label{tab3}%
\begin{tabular}{@{}lllll@{}}
\toprule
Attention & Robot-assisted Classroom  & Regular Classroom  \\
\midrule
SN001    & 4.99$\%$    & 35.15$\%$     \\
SN002    & 4.70$\%$   & 32.63$\%$     \\
SN003    & 25.20$\%$   & 26.67$\%$     \\
SN004    & 21.77$\%$   & 31.75$\%$     \\
SN005    & 12.15$\%$   & 37.27$\%$     \\
SN006    & 8.18$\%$       & 30.89$\%$     \\
\botrule
\end{tabular}
\end{table}

\begin{table}[b]
\caption{Proportion of Online Time and Time Spent “Looking at the Robot” by ASD Children in Robot-assisted Classroom}\label{tab4}%
\begin{tabular}{@{}lllll@{}}
\toprule
Attention & Online Time Proportion  & Looking at the NAO Robot\\
\midrule
SN001    & 76.48$\%$    & 65.81$\%$     \\
SN002    & 62.91$\%$   & 43.82$\%$     \\
SN003    & 65.38$\%$   & 30.93$\%$     \\
SN004    & 60.16$\%$   & 24.41$\%$     \\
SN005    & 65.72$\%$   & 37.92$\%$     \\
SN006    & 68.90$\%$   & 22.56$\%$     \\
\botrule
\end{tabular}
\end{table}

Specifically: 

SN001's attention in the robot-assisted classroom was significantly higher (76.48\%) compared to the regular classroom (47.26\%). He predominantly focused on the NAO robot, especially when it spoke or moved. Conversely, his attention to the teacher dramatically decreased in the robot-assisted classroom (4.99\%) compared to the regular setting (35.15\%). Notably, SN001 required less intervention for inappropriate behaviors in the robot-assisted setting, correlating with his increased focus on the robot.

SN002 displayed more attention in the robot-assisted classroom (62.91\%) than in the regular classroom (46.30\%), with substantial time spent observing the robot (43.82\%). Interestingly, his off-target behavior shifted from floor gazing in the regular classroom to looking at classmates in the robot-assisted setting, indicating the layout allowed better engagement.

SN003, the only participant with lower attention in the robot-assisted classroom (65.38\%) compared to the regular classroom (70.37\%), showed equal interest in the teacher across both settings but engaged distinctly during interactive sessions with the robot.

SN004's attention was similarly high in both settings, with slight variations in focus between the robot and the teacher. His behavior notably quieted down during the robot’s performances, suggesting effective engagement during these activities.

SN005 and SN006 both showed higher attention rates in the robot-assisted classroom compared to the regular classroom, with SN005 particularly focused during interactive demonstrations by the robot. Both students showed a pattern of frequent off-target behavior focused on classmates, indicating a common challenge across both classroom types.

\begin{table}[h]
\caption{Range, Mean, Median, and Standard Deviation of Attention Online Time Proportion in Two Classrooms}\label{tab5}%
\begin{tabular}{@{}lllll@{}}
\toprule
Attention & Regular Classroom  & Robot-assisted Classroom  \\
\midrule
Range    & 24.07$\%$    & 16.32$\%$     \\
Mean    & 57.01$\%$   & 66.59$\%$     \\
Median    & 57.26$\%$   & 65.55$\%$     \\
Standard Deviation    & 8.58$\%$   & 5.17$\%$     \\
\botrule
\end{tabular}
\end{table}

Table 5 lists four statistical measures of the proportion of online attention time of ASD children in both types of classrooms. The range (16.32\%) and standard deviation (5.17\%) of online attention time in robot-assisted classrooms were lower than those in regular classrooms (Range: 24.07\%, S.D.: 8.58), indicating a smaller fluctuation range and more stability in online attention time among ASD children in robot-assisted classroom. The mean (66.59\%) and median (65.55\%) of online attention time in robot-assisted classrooms were higher than in regular classrooms (Mean: 57.01\%, Median: 57.26\%), suggesting that the overall level of online attention time of ASD children in robot-assisted classroom was higher than in regular classroom.

\subsection{Comparative Analysis of Classroom-Related Communication for ASD Children in Both Types of Classrooms}\label{subsec4}

\begin{table}[h]
\caption{Comparison of Classroom-Related Communication in Two Classrooms}\label{tab6}%
\begin{tabular}{@{}lllll@{}}
\toprule
Communication & Regular Classroom  & Robot-assisted Classroom & Statement  \\
\midrule
SN001    & 22.87$\%$   & 42.02$\%$  & $\uparrow$   \\
SN002    & 10.92$\%$   & 19.11$\%$   & $\uparrow$  \\
SN003    & 36.21$\%$   & 48.29$\%$   & $\uparrow$  \\
SN004    & 18.70$\%$   & 29.48$\%$   & $\uparrow$  \\
SN005    & 41.75$\%$   & 42.94$\%$   & $\uparrow$  \\
SN006    & 37.37$\%$   & 56.25$\%$   & $\uparrow$  \\
\botrule
\end{tabular}
\end{table}

\begin{figure}[htbp]
\centering
\includegraphics[width=0.9\textwidth]{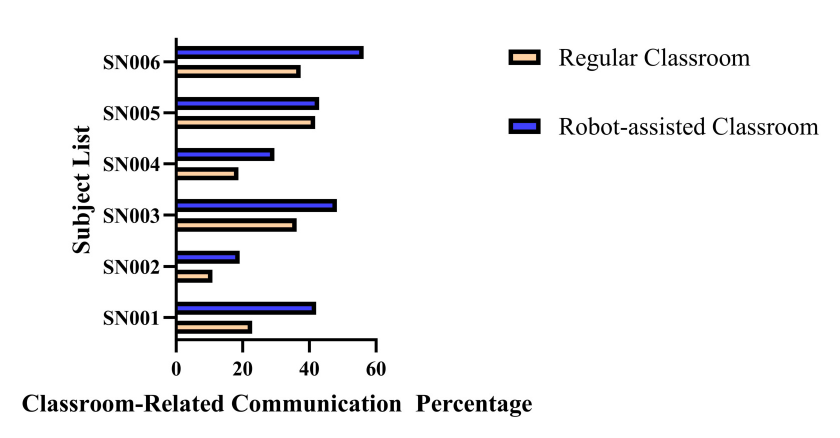}
\centering
\caption{Comparative Bar Chart of Subjects’Classroom-Related Communication in Two Classrooms}\label{fig8}
\end{figure}

Table 6 shows the data of classroom-related communication behaviors of ASD children in both types of classrooms. Although there are inevitable individual differences among ASD children, overall, the proportion of classroom-related communication in robot-assisted classrooms was greater than that in regular classrooms (see Fig 8). This suggests that children are more likely to participate in a classroom with NAO, and that a classroom with NAO has a higher effective response rate than a human teacher alone. 

\begin{table}[b]
\caption{Proportion of Classroom-Related Communication and Robot-Induced Communication in Robot-assisted Classroom}\label{tab7}%
\begin{tabular}{@{}lllll@{}}
\toprule
Communication & Classroom-Related Communication  & Robot-Induced Communication   \\
\midrule
SN001    & 42.02$\%$   & 21.01$\%$     \\
SN002    & 19.11$\%$   & 5.27$\%$     \\
SN003    & 48.29$\%$   & 15.61$\%$     \\
SN004    & 29.48$\%$   & 10.03$\%$     \\
SN005    & 42.94$\%$   & 9.60$\%$     \\
SN006    & 56.25$\%$   & 19.89$\%$     \\
\botrule
\end{tabular}
\end{table}

Specifically: 

SN001's classroom-related communication increased to 42.02\% from 22.87\% in the regular classroom. He actively engaged with the NAO robot, frequently exclaiming "wow" during activities and mimicking dance movements, with 21.01\% of his communications directly inspired by the robot (see Table 7).

SN002 showed improvement in the robot-assisted classroom with a participation rate rising from 10.92\% to 19.11\%. He interacted physically with the robot and was more involved in classroom activities like the picture-finding game, a noticeable change from his usual high-pitched screams in the regular classroom.

SN003 experienced an increase in communication from 36.21\% to 48.29\% when assisted by the robot, responding positively and frequently to NAO’s voice and movements. She actively participated during a photo-taking session, posing enthusiastically.

SN004’s participation also rose from 18.70\% to 29.48\% in the robot-assisted setting. He engaged more with the robot, initially hesitating but eventually responding to questions about his interests with the teacher's assistance.

SN005 maintained a consistent communication level, slightly increasing from 41.74\% to 42.94\%. Unlike in regular sessions where he was more passive, he interacted more dynamically with NAO, showing eagerness in hand-raising and responding to the robot’s queries.

SN006 significantly boosted his interaction to 56.25\% from 37.37\% in the robot-assisted classroom, showing a keen interest in the robot and participating eagerly in activities, even to the point of dancing and falling down in excitement.

\begin{table}[h]
\caption{Range, Mean, Median, and Standard Deviation of Classroom-Related Communication Proportion in Two Classrooms}\label{tab7}%
\begin{tabular}{@{}lllll@{}}
\toprule
Communication & Regular Classroom  & Robot-assisted Classroom   \\
\midrule
Range    & 30.83$\%$   & 37.14$\%$     \\
Mean     & 27.97$\%$   & 39.68$\%$     \\
Median   & 29.54$\%$   & 42.48$\%$     \\
Standard Deviation    & 11.17$\%$   & 12.19$\%$     \\
\botrule
\end{tabular}
\end{table}

Table 8 lists four statistical measures of the proportion of classroom-related communication produced by ASD children in both types of classrooms. All four categories of data in the robot-assisted classroom are greater than in the regular classroom, indicating that NAO, while attracting the attention of ASD children, also prompted them to engage in more classroom-related communication behaviors. However, it also reflects a slightly higher instability among the children in the robot-assisted classroom compared to the regular classroom, which may be related to the children’s initial encounter with the NAO robot.

\subsection{Comparative Analysis of Classroom Interaction Assessment of ASD Children in Both Types of Classrooms}\label{subsec4}

\begin{figure}[htbp]
\centering
\includegraphics[width=0.3\textwidth]{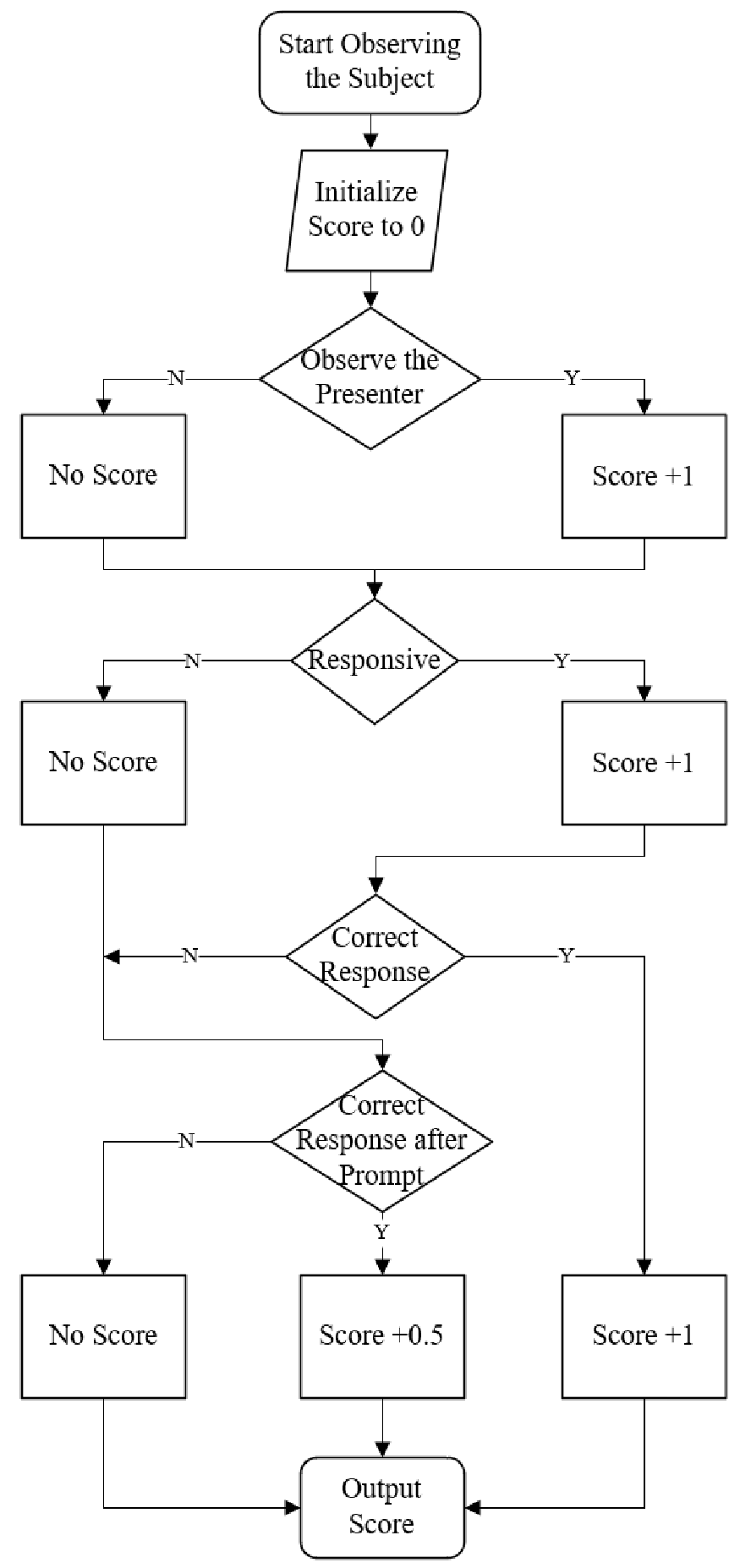}
\centering
\caption{Classroom Interaction Assessment Flowchart}\label{fig9}
\end{figure}

The classroom interaction assessment in this study utilized a comprehensive evaluation approach, scoring across three levels: the first level assesses whether children directed their gaze toward the activity presenter at the start of each activity; the second level evaluates if the children responded to the activity's initiation; the third level awards full credit if the child's response was correct. If incorrect, a prompt is given; a correct response post-prompt scores half, and no score is given for continued incorrect answers (see Fig 9). This method, a "relative comparison" rather than a single "absolute assessment," more accurately captures the students' overall performance. Standardized scores are used to vividly represent the assessment results of children with ASD.

\begin{table}[h]
\caption{Classroom Interaction Assessment Score in Two Classrooms}\label{tab7}%
\begin{tabular}{@{}lllll@{}}
\toprule
Assessment & Regular Classroom  &  Robot-assisted Classroom   \\
\midrule
SN001    & 52.1   & 76.7     \\
SN002    & 14.6   & 43.3     \\
SN003    & 75.0   & 74.4     \\
SN004    & 62.5   & 71.1     \\
SN005    & 66.7   & 70.0     \\
SN006    & 54.2   & 64.4     \\
Average  & 54.2   & 66.7     \\
\botrule
\end{tabular}
\end{table}

According to Table 9, the average score in the regular classroom was 54.2 points, while it rose to 66.7 points in the robot-assisted classroom. However, SN003 was an exception, scoring slightly lower in the robot-assisted setting compared to the regular classroom, as detailed in Fig 10. This section of the study highlighted several notable observations. For instance, during the circle-drawing activity, SN001 and SN003 persisted in using their left hands, even after prompts from NAO and the teacher to adjust. Additionally, in a group activity, SN006 mistakenly stood up when NAO called the first group, only sitting down after realizing the error and following the teacher's prompt. These instances suggest that while NAO effectively encourages participation, there is potential to further enhance the precision of children's responses and cognitive engagement in such interactions.

\begin{figure}[htbp]
\centering
\includegraphics[width=0.9\textwidth]{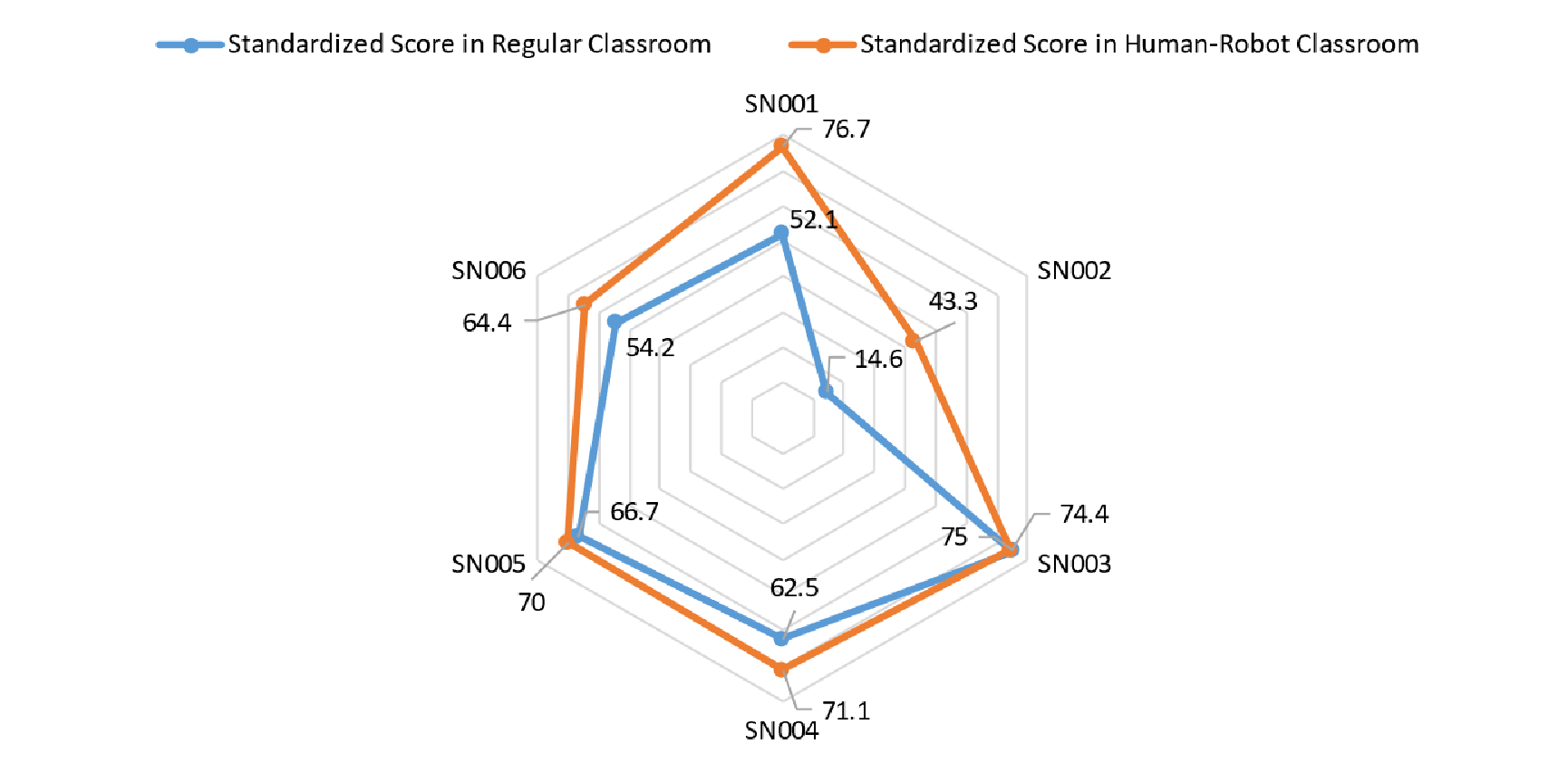}
\centering
\caption{Radar Chart Comparing Standardized Scores in Two Classrooms}\label{fig10}
\end{figure}

Looking at the comparative chart of standardized scores, it is evidence that the robot-assisted classroom positively promoted the children’s classroom interaction. Furthermore, the prompting phase after “response to questions” was used to try to cultivate the ASD children’s self-thinking ability and more social communication behaviors.

\subsection{Comparative Analysis of Emotional Proportion of ASD Children in Both Types of Classrooms}\label{subsec4}

\begin{table}[h]
\caption{Emotional Proportion of ASD Children in Two Classrooms}\label{tab10}%
\begin{tabular}{@{}lllllll@{}}
\toprule
Emotion & \multicolumn{3}{|c|}{Regular Classroom}  & \multicolumn{3}{|c|}{Robot-assisted Classroom}   \\

\midrule
         & Positive   & Negative   & Neutral   & Postive & Negative  & Neutral   \\
SN001    & 45.19   & 22.92   & 31.89   & 62.17 & N/A  & 37.83\\
SN002    & 2.68   & 62.75   & 34.57   & 15.78 & 33.33  & 50.89\\
SN003    & 26.80   &10.74   & 62.46   & 46.91 & 8.96  & 44.13\\
SN004    & 12.00   & 35.32   & 52.68   & 30.93 & 35.19  & 33.88\\
SN005    & 34.65   & 23.67   & 41.68   & 42.08 & 16.88  & 41.04\\
SN006    & 4.95   & 14.94   & 80.11   & 7.19 & 12.48  & 80.33\\
\botrule
\end{tabular}
\end{table}

Table 10 records the three categories of emotional proportions of ASD children in both types of classrooms. Specifically, SN001 often smiled at NAO in the robot-assisted classroom and sometimes applauded NAO. During the entire robot-assisted course, no significant negative emotions were observed in SN001 (marked as N/A in the table), but instances of whimpering and other classroom disruptions occurred in the regular classroom. SN002 often appeared disengaged in the regular classroom, sometimes climbing up to tables and shaking arms, or lying on multiple chairs. Despite being briefly calmed by the teacher’s prompts, he suddenly started wandering around again, engaging in off-topic activities. However, in the robot-assisted classroom, his negative behaviors significantly decreased, with no climbing on tables or lying on chairs, although he did leave his seat twice to touch the robot. NAO’s speech and movements attracted him and involved him in some classroom activities. SN003 paid great attention to NAO’s voice, pulling SN004’s arm to direct his attention to the robot whenever NAO spoke, and she showed smiley face. She appeared shy in response to NAO’s praise and compliments. SN004 experienced a wide range of emotions throughout the class, from crying and resisting upon entering the classroom to being reluctant to leave at the end. Although attracted by NAO’s appearance and voice, his off-topic talking, emotional stomping, and body shaking did not decrease. SN005 maintained good emotional control in both classrooms, showing interest in NAO, and frequently interacting with a smiley face. He laughed when NAO thanked and praised him. SN006 rarely showed emotions in the regular classroom, but in the robot-assisted classroom, he displayed strong curiosity towards NAO, focused on NAO’s talent show with applause, and remained expressionless for most of the time.

\begin{figure}[h]
\centering
\includegraphics[width=0.9\textwidth]{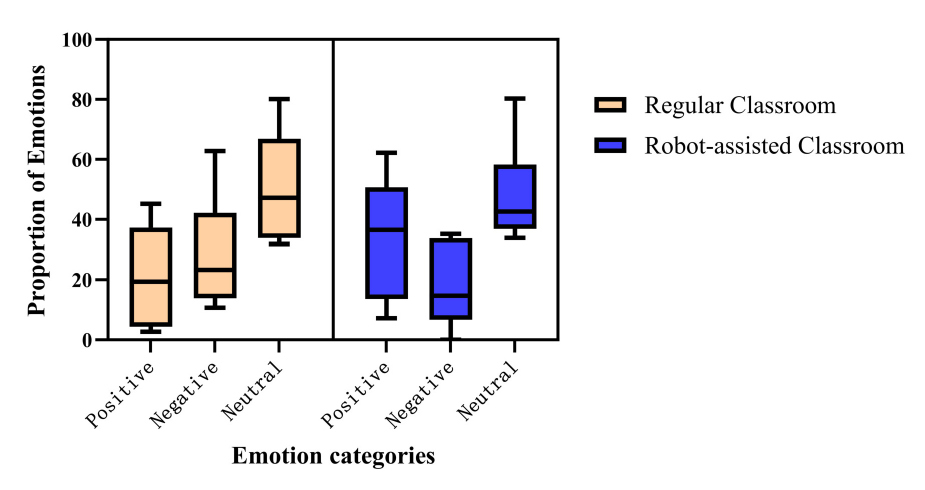}
\centering
\caption{Boxplot of Emotional Proportion of ASD Children in Two Classrooms}\label{fig11}
\end{figure}

The boxplot compares the emotional proportions in both types of classrooms (see Fig 11) visually reveals that the robot-assisted classroom had higher extremes and medians for positive emotion and lower extremes and medians for negative emotion compared to the regular classroom. 

 \section{Conclusion}\label{sec5}
The findings of this research lend partial support to our initial hypotheses, elucidating the impacts of robot-assisted classrooms on ASD children. Hypothesis 1) was partially validated: apart from subject SN003, online attention in robot-assisted classrooms was generally higher and more stable than in regular classrooms. Hypothesis 2) was fully confirmed as interactions with the NAO robot significantly boosted classroom-related communication among ASD students, despite the persistence of off-topic discussions. Similarly, Hypothesis 4) was validated, with students displaying higher positive emotions and fewer negative emotions in the robot-assisted environment compared to the traditional setting. However, for Hypothesis 3), except for SN003, only other students showed improved standardized scores in the robot-assisted classrooms.

The study underscores that the presence of the NAO robot in group teaching settings captivates ASD children, fostering greater engagement and enhancing educational outcomes. These children not only focused better but also participated more in classroom communication, scored higher in activities, and demonstrated a deeper connection with NAO than those in regular classrooms. Moreover, the emotional responses elicited by NAO's praise and motivational speech were a notable and unexpected discovery, emphasizing the potential of such technologies to enrich the educational experience for ASD children.

Overall, our experiment assessed the efficacy of robot-assisted learning across four key metrics—attention, communication, academic performance, and emotional well-being—and observed significant enhancements in all areas. This study forms part of our ongoing research into human-robot interaction in special education. We are currently advancing our investigations with the NAO robot, aiming to refine our experimental approaches and teaching strategies to better serve and integrate ASD children into societal frameworks.

\bmhead{Acknowledgements}

This research was organized by Changshu Institute of Technology and Changshu Special Education School and this study was supported by the MOE of Humanities and Social Sciences Project Fund of the Republic of China (21YJCZH029) and SIP Autism Spectrum Disorder Emotional Recognition AI Technology Innovation Platform (YZCXPT2022107, YZCXPT2022106 and YZCXPT2023103). Special thanks to Fangxian Zhang, Jing Zhang, Xinru Zhang, Aowei Shen, and all who have contributed to this work.


\end{document}